\PassOptionsToPackage{margin=2cm}{geometry}
\documentclass[sn-nature]{sn-jnl}

\usepackage{graphicx}%
\usepackage{multirow}%
\usepackage{amsmath,amssymb,amsfonts}%
\usepackage{amsthm}%
\usepackage{mathrsfs}%
\usepackage[numbers]{natbib} 
\usepackage[title]{appendix}%
\usepackage{xcolor}%
\usepackage{textcomp}%
\usepackage{manyfoot}%
\usepackage{booktabs}%
\usepackage{algorithm}%
\usepackage{algorithmicx}%
\usepackage{algpseudocode}%
\usepackage{listings}%
\usepackage{comment}%
\usepackage{soul}%
\usepackage{amsfonts}%
\usepackage{booktabs}%
\usepackage{siunitx}%
\usepackage{multirow}%
\usepackage{float}%
\usepackage{tikz}
\def\checkmark{\tikz\fill[scale=0.4](0,.35) -- (.25,0) -- (1,.7) -- (.25,.15) -- cycle;} 
\usepackage{flafter}

\theoremstyle{thmstyleone}%
\theoremstyle{thmstyletwo}%

\theoremstyle{thmstylethree}%

\begin{document}

\title[Interoperability definition]{Defining Interoperability: a universal standard}


\author*[1]{\fnm{Giada} \sur{Lalli}}\email{giada.lalli@kuleuven.be}

\affil*[1]{\orgdiv{Department of Human Genetics}, \orgname{KU Leuven}, \orgaddress{\street{Herestraat 49}, \city{Leuven}, \postcode{3000}, \state{Flemish Brabant}, \country{Belgium}}}


\abstract{Interoperability is crucial for modern scientific advancement, yet its fragmented definitions across domains hinder researchers' ability to effectively reap the rewards. This paper proposes a new, universal definition by tracing the evolution of interoperability and identifying challenges posed by varying definitions. This definition addresses these inconsistencies, offering a robust solution applicable across diverse fields. Adopting this unified approach will enhance global collaboration and drive innovation by removing obstacles to interoperability posed by conflicting or incomplete definitions.}

\keywords{interoperability, standardization, cross-domain}



\maketitle
Interoperability is vital for modern science and technology, especially in life sciences, where seamless data exchange and collaboration are crucial for innovation and efficiency~\cite{gurdur2018systematic}.
Despite its critical importance, interoperability has been a longstanding challenge. Since its inception in the late 20th century \cite{ford2007survey}, the concept has evolved significantly, driven by the increasing complexity of information systems and the need for more sophisticated interactions.

However, this evolution has also led to fragmentation. With over 117 distinct definitions~\cite{maciel2024systems} of interoperability documented across various domains, the lack of a unified understanding has created barriers to effective innovation and efficiency. Different industries and application domains, from healthcare to logistics, interpret interoperability in ways that suit their immediate needs, leading to inconsistencies that hamper effective working between users of different definitions \cite{maciel2024systems}.
To address the fragmentation caused by divergent definitions, we must first understand the historical evolution of interoperability.

\section*{History of Interoperability}
Interoperability, defined by the U.S. Department of Defense (DoD) in 1977 \cite{ford2007survey} as
\begin{quote}
\textit{the ability of systems, units, or forces to provide services to and accept services from other systems, units, or forces and to use the services so exchanged to enable them to operate effectively together.}
\end{quote}

This definition addressed critical military challenges. Military communication systems lacked compatibility, hindering joint operations. The concept of interoperability emphasized the need for standardized protocols to ensure seamless technology integration. 
Since then, interoperability has significantly evolved, both in its numerous definitions and the associated distinct types~\cite{ford2007survey, maciel2024systems}. 
By the 1980s, the DoD broadened the concept to include \textit{electronic} and \textit{logistic} interoperability, emphasizing compatible interfaces and components. 
In the 1990s, definitions widened to include \textit{technical} interoperability, focusing on dynamic information exchange and operational integration. 

The Institute of Electrical and Electronics Engineers (IEEE) significantly contributed to this evolution, emphasizing basic information exchange among systems from 1990 onward with the publication of the Standard Computer Dictionary~\cite{182763}. By 2000, the IEEE's definitions included the ability of equipment to work together in networks, stressing the importance of standards for interoperability. Concurrently, other definitions highlighted the effort required to couple systems (1980, \cite{maccall1980software}), the quality of information exchange (1987, \cite{poppel1987information}), and the ability of large-scale distributed systems to exchange services and data (1995, \cite{heiler1995semantic}).

In the 2000s, the concept of interoperability evolved from a containerised focus to broader applications, incorporating semantic, structural, and syntactic aspects to address the growing complexity of information systems. The European Interoperability Framework (EIF, \cite{doi/10.2799/17759}) expanded this understanding by emphasizing seamless data, business process, and organizational interactions. Similarly, projects like IDEAS (2005, \cite{blanc2005interoperability}) highlighted the need for interoperability across data, application, and business process levels in enterprise software. 

Commonalities among these definitions include a consistent emphasis on information exchange and utilization, the importance of technical compatibility, and the role of standards. Differences emerge in the specific contexts and domains addressed: the DoD focuses on operational and logistic aspects, IEEE emphasizes technical standards and heterogeneous environments, and other definitions highlight semantic integration and enterprise applications.

While these developments mark significant progress, they also indicate the ongoing challenges in achieving true interoperability. The proliferation of definitions and standards across sectors has led to inconsistencies and gaps in implementation, particularly in complex domains like healthcare. These challenges highlight the urgent need for a comprehensive, unified approach to interoperability. 


\section*{Proposed definition}
The proposed new definition articulates in two propositions: 
\subsubsection*{Proposition 1: a concise definition}
\begin{quote}
Interoperability is the capacity of diverse systems, units, or components to seamlessly exchange information, services, and data, utilising this exchanged content in a meaningful, effective, and efficient manner. 
\end{quote}

\subsubsection*{Proposition 2: an extended definition}
\begin{quote}
Interoperability encompasses several dimensions:

\begin{itemize}
    \item \textbf{Technical Interoperability}: refers to the ability of diverse systems, networks, and platforms to communicate and exchange information seamlessly. This involves facilitating dynamic, interactive data and service exchanges without the need for translation or middleware, ensuring that systems speak the same language. It includes the integration of blockchain networks, cloud services, and interconnected networks, all of which must operate together efficiently to maintain expected quality levels. 
    \item \textbf{Pragmatic Interoperability}: ensures that the messages exchanged between systems achieve their intended effects and are correctly understood by collaborating systems. This requires systems to be syntactically and semantically aligned and dynamically adaptable to changes in context over time. 
    \item \textbf{Semantic Interoperability}:  enables systems to exchange information with a shared understanding of meaning, context, and purpose. It ensures that data, concepts, and knowledge are interpreted consistently across different systems, cultures, and domains. This includes the alignment of conceptual models, the shared understanding of data within ecosystems, and the ability to transfer knowledge across heterogeneous environments. 
    \item \textbf{Syntactic Interoperability}: ensures that data exchanged between systems follows a compatible format, allowing it to be processed and understood without requiring additional transformation. This involves defining clear data structures and formats that facilitate seamless communication between diverse systems—also known as \textit{structural} interoperability.
    \item \textbf{Legal Interoperability}: Ensuring effective collaboration across organisations with different legal frameworks. It involves adapting legislation to support cross-border services and establishing clear agreements to address legal differences, promoting flexibility and scalability in diverse legal environments.
    \item \textbf{Organisational Interoperability}: involves the alignment and integration of business processes, policies, and information exchanges across different entities to meet user needs effectively. It encompasses the ability of enterprises to cooperate with partners, conduct IT-supported business relationships, and align their processes seamlessly, enabling businesses to achieve their objectives automatically and efficiently.
    \item \textbf{Operational Interoperability}: enables systems, units, or organizations to provide and access services from one another, ensuring effective and coordinated operations. This type of interoperability is crucial in contexts like multinational military coalitions, where diverse forces must work together cohesively in real time. 
    \item \textbf{Enterprise Interoperability}: the ability of an enterprise—a company or other large organization—to functionally link activities, such as product design, supply chains, and manufacturing, efficiently and competitively. 
    \item \textbf{Constructive Interoperability}: using common architecture, standards, data specifications, and communication protocols to build and maintain interoperable systems. 
    \item \textbf{Programmatic Interoperability}: managing the interaction of one program within the context of another, ensuring coherent program management. 
    \item \textbf{Electronic Interoperability}: facilitates the effective and secure exchange of information among devices, systems, and platforms through common interface characteristics. This includes the integration of Internet of Things (IoT) devices, ensuring they can work together within the same ecosystem, and the ability of electronic devices to communicate and control one another seamlessly.
    \item \textbf{Logistic Interoperability}: exchanging assemblies, components, spares, or repair parts, sometimes accepting performance degradation if operationally acceptable. 
    \item \textbf{Data Interoperability}: enables the seamless exchange, interpretation, and integration of data across different systems and organizational boundaries. It ensures that data can be merged, aggregated, reused, and processed in meaningful ways, with shared expectations for content, context, and quality. This interoperability is critical for supporting clarity, scalability, and innovation across various domains.
\end{itemize}
\end{quote}
This novel definition provides a robust foundation for addressing the core challenges of fragmentation, lack of consensus, and limited applicability, offering the scientific community a unified and adaptable framework. By encompassing a wide spectrum of interoperability dimensions—from technical to legal—it aims to resolve inconsistencies and gaps in current practices. For instance, semantic interoperability ensures that exchanged information is interpreted consistently across systems, reducing the misunderstandings that frequently plague current implementations. Additionally, operational and programmatic interoperability facilitate cross-domain collaboration, ensuring that sectors ranging from healthcare to logistics can communicate and operate effectively within a cohesive framework. 

Our proposed unified definition consolidates the most stable and widely accepted types of interoperability, while clarifying and addressing previous misunderstandings \cite{maciel2024systems}. This approach simplifies the interoperability landscape, offering a clear, comprehensive foundation that supports the seamless integration of new technologies and domains, ensuring future advancements can be adopted effectively 

The adoption of these comprehensive definitions will have a profound impact across multiple fields. In healthcare, a universal understanding of interoperability will streamline patient data exchange, reduce errors and improve outcomes. In logistics, it will enhance coordination between supply chain systems, boosting efficiency and cutting costs. In information technology, it will provide a solid framework for integrating emerging technologies like AI and IoT into existing systems, fostering innovation and ensuring long-term scalability.

\section*{Evaluation of Interoperability Definitions}

In this paper, I evaluate definitions of interoperability on five key criteria:
\begin{description}
    \item[Flexibility] The definition is adaptable to both future advancements and to different contexts than that in which it was originally proposed.
    \item[Clarity \& Conciseness] The definition uses simple language that is easy to understand.
    \item[Measurability] The definition allows for the assessment, bench-marking, and otherwise evaluation of interoperability efforts.
    \item[Scalability] The definition is applicable across application scales: from small-scale local applications to very large-scale global applications.
    \item[Language \& Standards] The definition establishes a common language and set of standards that can be universally applied.
\end{description}

\noindent These criteria are marked as either present (\checkmark) or absent in Table~\ref{table:evaluation}. DoD definitions are clear, measurable, and use common language. The Creps et al. (2008)~\cite{creps2008systems} definition, while clear, concise, and uses common language, is limited to computer systems. The definition proposed by Asuncion \& van Sinderen (2010)~\cite{10.1007/978-3-642-15509-3_15}, though flexible and measurable, lacks scalability. The National Interoperability Framework Observatory (NIFO) definition proposed by Kalogirou~\cite{10.1007/978-3-030-13693-2_30} is flexible and scalable but is difficult to fully comprehend. Finally, the definition proposed here addresses all criteria and so is a suitable universal definition by these criteria.

\begin{table}[ht]
\centering
\caption{Evaluation of a representative selection of Interoperability Definitions.\label{table:evaluation}}
\begin{tabular}{p{3.5cm}|l|ccccc}
\hline
\multicolumn{1}{c|}{\multirow{2}{*}{\centering Source}} & \multicolumn{1}{c|}{\multirow{2}{*}{\centering Definition}} & \multicolumn{5}{c}{Criteria} \\ 
\multicolumn{1}{c|}{}                        & \multicolumn{1}{c|}{}  & \rotatebox{90}{Flexibility} & \rotatebox{90}{Clarity and Conciseness} & \rotatebox{90}{Measurability} & \rotatebox{90}{Scalability} & \rotatebox{90}{Language and Standards} \\ \hline
DoD (1977),~\cite{ford2007survey} & \begin{tabular}[c]{@{}l@{}}The ability of systems, units, or forces to provide \\ services to and accept services from other systems, \\ units, or forces and to use the services so exchanged \\ to enable them to operate effectively together.\end{tabular} &  &  &  &  & \checkmark \\ \hline
DoD (1997),~\cite{ford2007survey} & \begin{tabular}[c]{@{}l@{}}Technical interoperability: The condition achieved \\ among communications-electronics systems or \\ items of communications-electronics equipment \\ when information or services can be exchanged directly\\ and satisfactorily between them and/or their users. \\ The degree of interoperability should be defined when \\ referring to specific cases.\end{tabular} &  & \checkmark & \checkmark &  &  \\ \hline
Creps et al. (2008),~\cite{creps2008systems} & \begin{tabular}[c]{@{}l@{}}Semantic interoperability is the ability of computer \\ systems to exchange data with unambiguous, shared \\ meaning. Semantic interoperability is a requirement \\ to enable machine computable logic, inferencing, \\ knowledge discovery, and data federation between\\ information systems.\end{tabular} &  & \checkmark &  &  & \checkmark \\ \hline
Asuncion and van Sinderen (2010),~\cite{10.1007/978-3-642-15509-3_15} & \begin{tabular}[c]{@{}l@{}}To ensure pragmatic interoperability, message sent \\ by a system causes the effect intended by that system; \\ i.e., the intended effect of the message is understood \\ by the collaborating systems. Pragmatic interoperability \\ can only be achieved if systems are also syntactically \\ and semantically interoperable.\end{tabular} & \checkmark &  & \checkmark &  &  \\ \hline
Asuncion and van Sinderen (2010),~\cite{10.1007/978-3-642-15509-3_15} & \begin{tabular}[c]{@{}l@{}}To ensure syntactic interoperability, collaborating \\ systems should have a compatible way of structuring \\ data during exchange; i.e., the manner in which data \\ is codified using a grammar or vocabulary is \\ compatible.\end{tabular} &  & \checkmark &  &  & \checkmark \\ \hline
NIFO (2019),~\cite{10.1007/978-3-030-13693-2_30} & \begin{tabular}[c]{@{}l@{}}Legal interoperability is about ensuring that organisations \\ operating under different legal frameworks, policies and \\ strategies are able to work together. This might require \\ that legislation does not block the establishment of European \\ public services within and between Member States and that \\ there are clear agreements about how to deal with differences \\ in legislation across borders, including the option of \\ putting in place new legislation.\end{tabular} & \checkmark &  &  & \checkmark & \checkmark \\ \hline
\textbf{Proposed definition} & \textbf{Proposition 1 and 2 above} & \checkmark & \checkmark & \checkmark & \checkmark & \checkmark \\ \hline
\end{tabular}%
\end{table}



\subsection*{Case Study: healthcare interoperability}
The extensive challenge of interoperability is easily illustrated within the healthcare sector, where the intricate nature of data and the heterogeneity of systems have long hindered seamless integration \cite{braunstein2018health, bhartiya2013exploring}. 

The healthcare ecosystem encompasses a diverse array of professionals—such as doctors, nurses, and pharmacists—as well as various computing systems and specialized domains, all of which generate and process a wide range of data, including clinical records and laboratory results. The absence of standardized definitions and protocols for data exchange has precipitated significant interoperability issues~\cite{iroju2013interoperability}. For instance, inconsistencies in how electronic health records (EHRs) define ``medical records" can result in fragmented data exchanges, severely impacting patient care \cite{janett2020electronic}. Such discrepancies can manifest as incomplete patient histories, diagnostic errors, medication mishaps, inconsistent treatment plans, fragmented care coordination, and flawed data reporting. These issues collectively and systematically undermine the quality of care by depriving healthcare providers of essential, comprehensive information. 

While standards such as HL7~\cite{hl7_website, benson2016principles} and DICOM~\cite{mustra2008overview} have been established, they have not resolved the underlying fragmentation, characterised by persistent inconsistencies and divergent interpretations. This fragmentation can lead to critical information being lost or misinterpreted, particularly during transitions of care between providers \cite{lapke2017disconnect, tu2015data}.

The lack of a universal foundation in the form of a definition of interoperability that is \textit{flexible}, \textit{clear and concise}, \textit{measurable}, \textit{scalable}, and \textit{standardised} leads to persistent communication inconsistencies, resulting in inefficiencies and potential risks to patient care. This underscores the urgent need for a comprehensive and universally applicable definition of interoperability.

\section*{Conclusion}
A single, universally accepted definition of interoperability is essential for unifying efforts across diverse domains and improving outcomes in critical areas like healthcare. To overcome the challenges of fragmented systems, a consistent and comprehensive approach is needed. This paper proposes a universal definition of interoperability that meets all key criteria, offering a clear path forward for enhancing collaboration and effectiveness across fields. 


\section*{Declarations}
\begin{itemize}
\item \textit{Funding}: Not applicable.
\item \textit{Competing interests}: The author declares no competing interests. 
\end{itemize}

\section*{Acknowledgements}
The author is deeply grateful to Dr. J. Collier for his support in writing and revising the text: without his help, this result would not have been possible.
\noindent

\bigskip


\bibliography{bibliography}

\end{document}